\documentclass{iau} 
\usepackage{graphicx}

\title[IAUS330.~~Abundance ratios and ages] %% give here short title %%
{Abundance ratios \& ages of stellar \\ populations in HARPS-GTO sample}

\author[E. Delgado Mena \etal\ ]   %% give here short author list %%
{E. Delgado Mena$^1$, M. Tsantaki$^2$, V. Zh. Adibekyan$^1$, S. G. Sousa$^{(1,3)}$, N.~C.~Santos$^{(1,3)}$, J.~I.~Gonz\'alez Hern\'andez$^{(4,5)}$
      \and G. Israelian$^{(4,5)}$ 
}

\affiliation{

$^1$Instituto de Astrof\'isica e Ci\^encias do Espa\c{c}o, \\Universidade do Porto, CAUP, Rua das Estrelas, PT4150-762 Porto, Portugal. \\ email: {\tt elisa.delgado@astro.up.pt} \\[\affilskip]
$^2$Instituto de Radioastronom\'ia y Astrof\'isica, IRyA, UNAM, \\Campus Morelia, A.P. 3-72, C.P. 58089, Michoac\'an, Mexico \\[\affilskip]     
$^3$Departamento de F\'isica e Astronom\'ia, Faculdade de Ci\^encias, U. Porto, Portugal \\[\affilskip]
$^4$Instituto de Astrof\'{\i}sica de Canarias,C/ Via Lactea, s/n, 38205, La Laguna, Tenerife, Spain \\[\affilskip]
$^5$Departamento de Astrof\'isica, Universidad de La Laguna, 38206 La Laguna, Tenerife, Spain \\[\affilskip]
}

\pubyear{2017}
\volume{330}  %% insert here IAU Symposium No.
\jname{Astrometry ans Astrophysics in the Gaia Sky}
\editors{A.C. Editor, B.D. Editor \& C.E. Editor, eds.}
\begin{document}

\maketitle

\begin{abstract}

In this work we present chemical abundances of heavy elements (Z$>$28) for a homogeneous sample of 1059 stars 
from HARPS planet search program. We also derive ages using parallaxes from Hipparcos and Gaia DR1 
to compare the results. We study the [X/Fe] ratios for different populations and compare them 
with models of Galactic chemical evolution. We find that thick disk stars are 
chemically disjunt for Zn and Eu. Moreover, the high-alpha metal-rich population presents an interesting 
behaviour, with clear overabundances of Cu and Zn and lower abundances of Y and Ba with respect to thin disk stars.
Several abundance ratios present a significant correlation with age for chemically separated thin disk stars 
(regardless of their metallicity) but thick disk stars do not present that behaviour. 
Moreover, at supersolar metallicities the trends with age tend to be weaker for several elements.

\end{abstract}

\firstsection % if your document starts with a section,
\section{Introduction}

In the era of large spectroscopic surveys such as APOGEE, Gaia-ESO Survey or RAVE, among others, 
the contribution of smaller samples with high-resolution and high quality spectra is of great 
importance to understand the Galactic Chemical Evolution (GCE). In this work we have derived abundances
for Cu, Zn, Sr, Y, Zr, Ba, Ce, Nd and Eu (\cite[Delgado Mena et al. 2017]{delgadomena17}) for 1111 stars 
within the volume-limited HARPS-GTO planet search sample in order to complement our previous works for
light elements (\cite[Delgado Mena et al. 2014]{delgadomena14}, \cite[Delgado Mena et al. 2015]{delgadomena15},
\cite[Su\'arez Andr\'es et al. 2016]{suarezandres16}, \cite[Bertr\'an de Lis et al. 2015]{bertrandelis15}), 
$\alpha$- elements and Fe-peak elements (\cite[Adibekyan et al. 2012]{adibekyan12}). The main purpose of 
this work is to evaluate the GCE evolution of those heavier elements and the dependence on 
stellar ages of different abundance ratios.

\begin{figure}[t!]
% \vspace*{-2.0 cm}
\begin{center}
 \includegraphics[width=13.5cm]{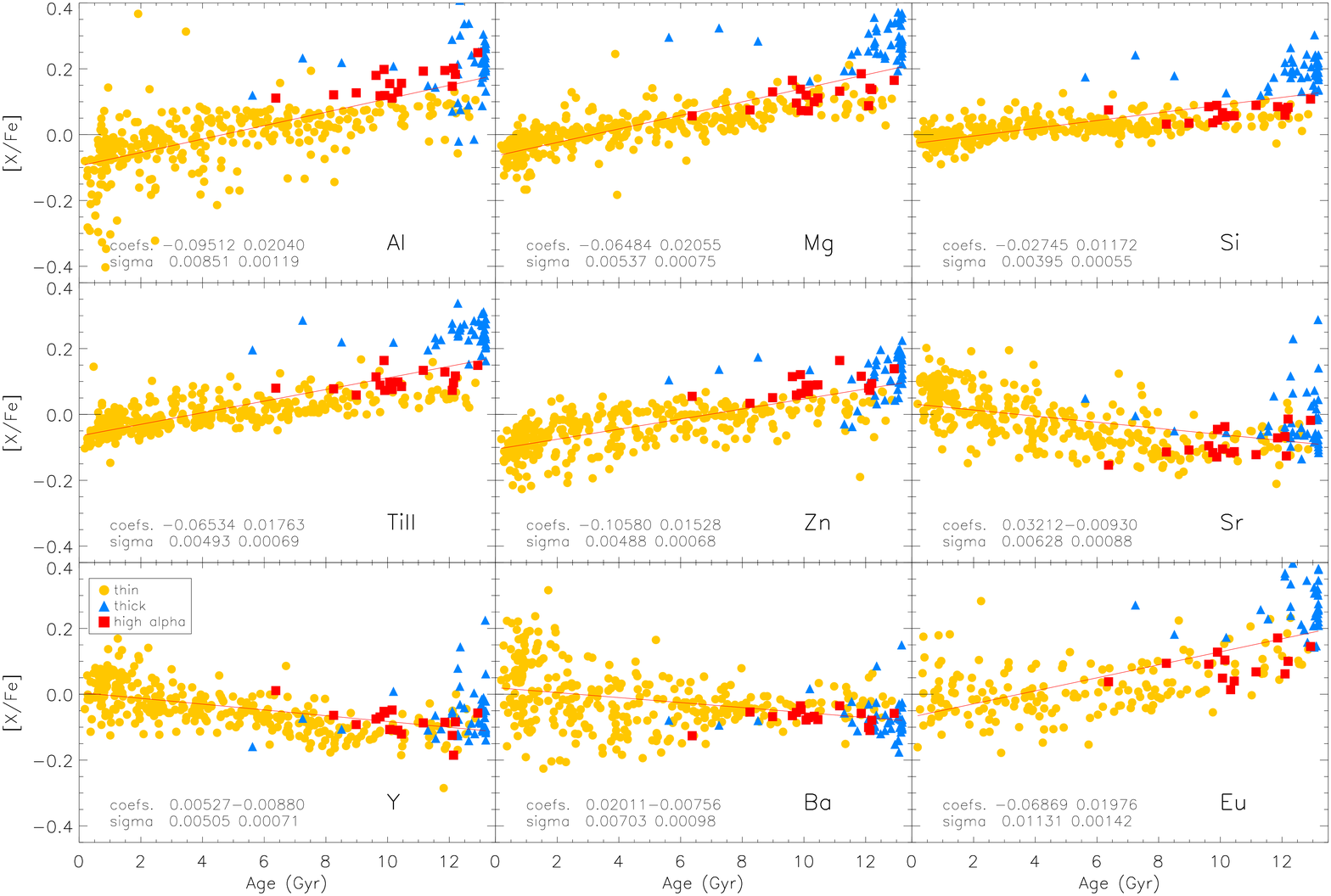} 
% \vspace*{-1.0 cm}
 \caption{General [X/Fe] ratios as a function of age for the reduced sample with reliable ages. 
 Thin disk stars, thick disk stars and \textit{h$\alpha$mr} are depicted with dots. triangles and 
 squares, respectively. The linear fit to all the stars just serves as eye guiding.}
   \label{fig1}
\end{center}
\end{figure}

\section{Stellar ages}
We derive the masses, radii and ages with the PARAM v1.3 tool\footnote{http://stev.oapd.inaf.it/cgi-bin/param} 
using the PARSEC isochrones (\cite[Bressan et al. 2012]{bressan12}) with our values for Teff 
and [Fe/H], the V magnitudes from the main Hipparcos catalogue (\cite[Perryman et al. 1997)]{perryman97}) 
and the parallaxes from the Hipparcos new reduction (\cite[van Leeuwen 2007]{vanleeuwen07}) 
or from the first release (DR1) of Gaia (\cite[Lindegren et al. 2016]{lindegren16}). We note that we added a 
systematic error of 0.3 mas to the formal error of the Gaia DR1 parallaxes as recommended by 
the Gaia collaboration. Meanwhile Hipparcos provides parallaxes for 1051 out of the 1059 
stars within our sample, only 923 stars have parallaxes in GAIA DR1. Moreover, there are significant
differences in many cases, leading to non-negligible differences in age. In order to have a sample with
ages as reliable as possible we decided to select the Hipparcos ages with a difference less than 1 Gyr with 
respect to the ages derived with GAIA parallaxes and with an error in age lower than 2 Gyr. This final sample
is composed by 377 stars belonging to the thin disk, thick disk and high-$\alpha$ metal-rich stars 
(hereafter \textit{h$\alpha$mr}, a population with high $\alpha$ abundances at [Fe/H]\,$>$\,-0.2\,dex discovered
by \cite[Adibekyan et al. 2011]{adibekyan11}).

\section{Abundance ratios vs age}
\begin{figure}[t!]
% \vspace*{-2.0 cm}
\begin{center}
 \includegraphics[width=6.6cm]{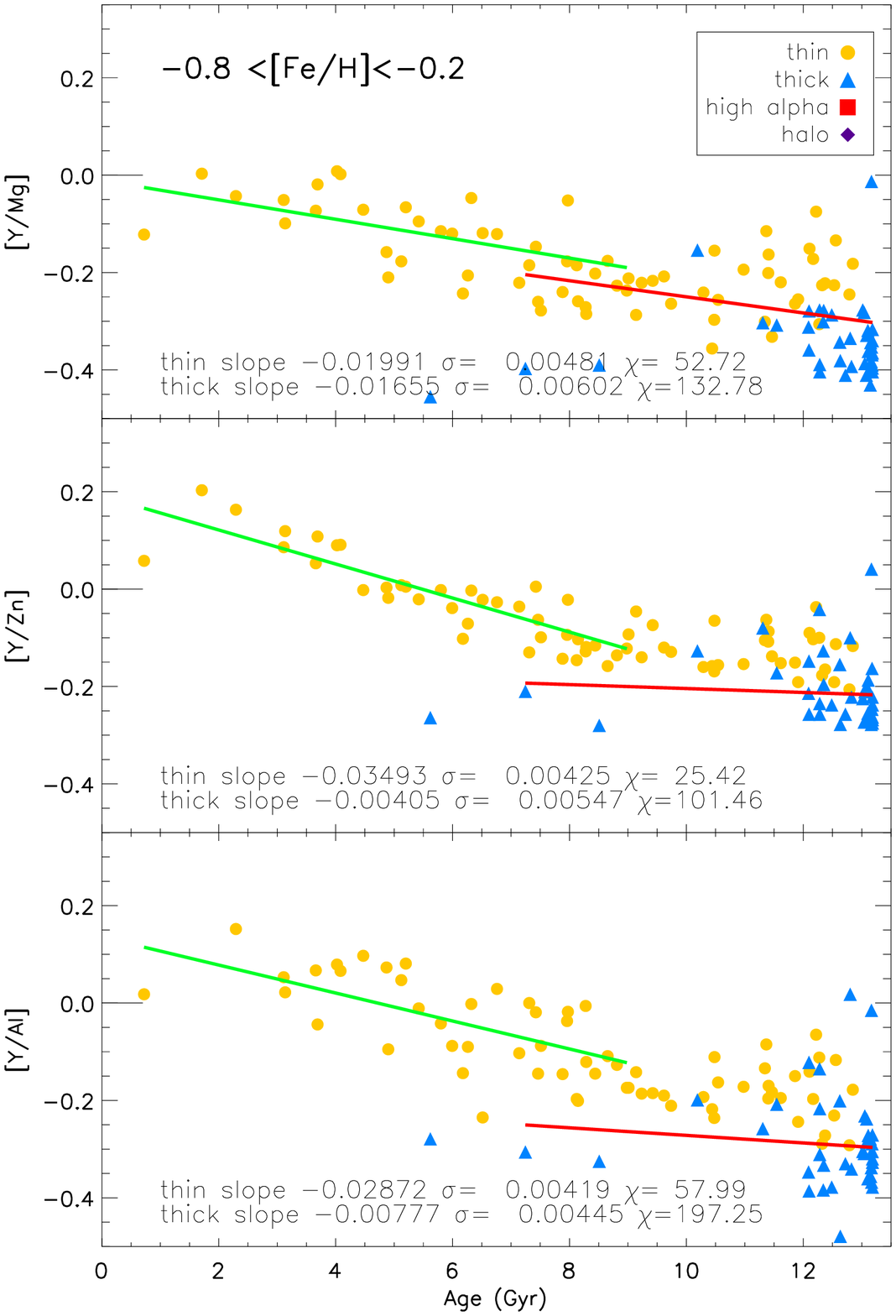}
 \includegraphics[width=6.6cm]{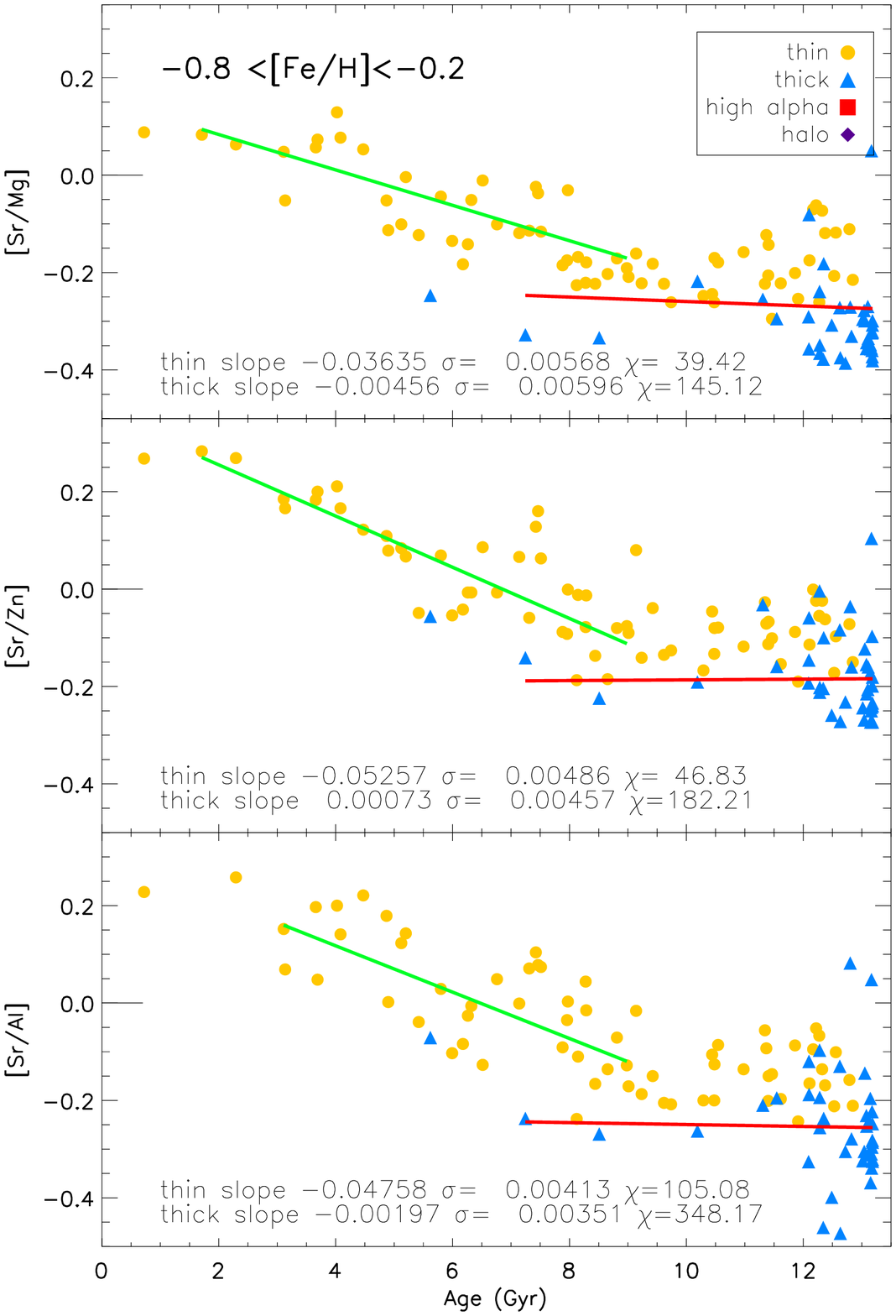}
 \caption{[Y/Mg],[Y/Zn],[Y/Al] and [Sr/Mg],[Sr/Zn],[Sr/Al] for -0.8$<$[Fe/H]$<$-0.2. Symbols as in Fig 1.}
   \label{fig2}
\end{center}
\end{figure}

\begin{figure}[t!]
% \vspace*{-2.0 cm}
\begin{center}
 \includegraphics[width=6.6cm]{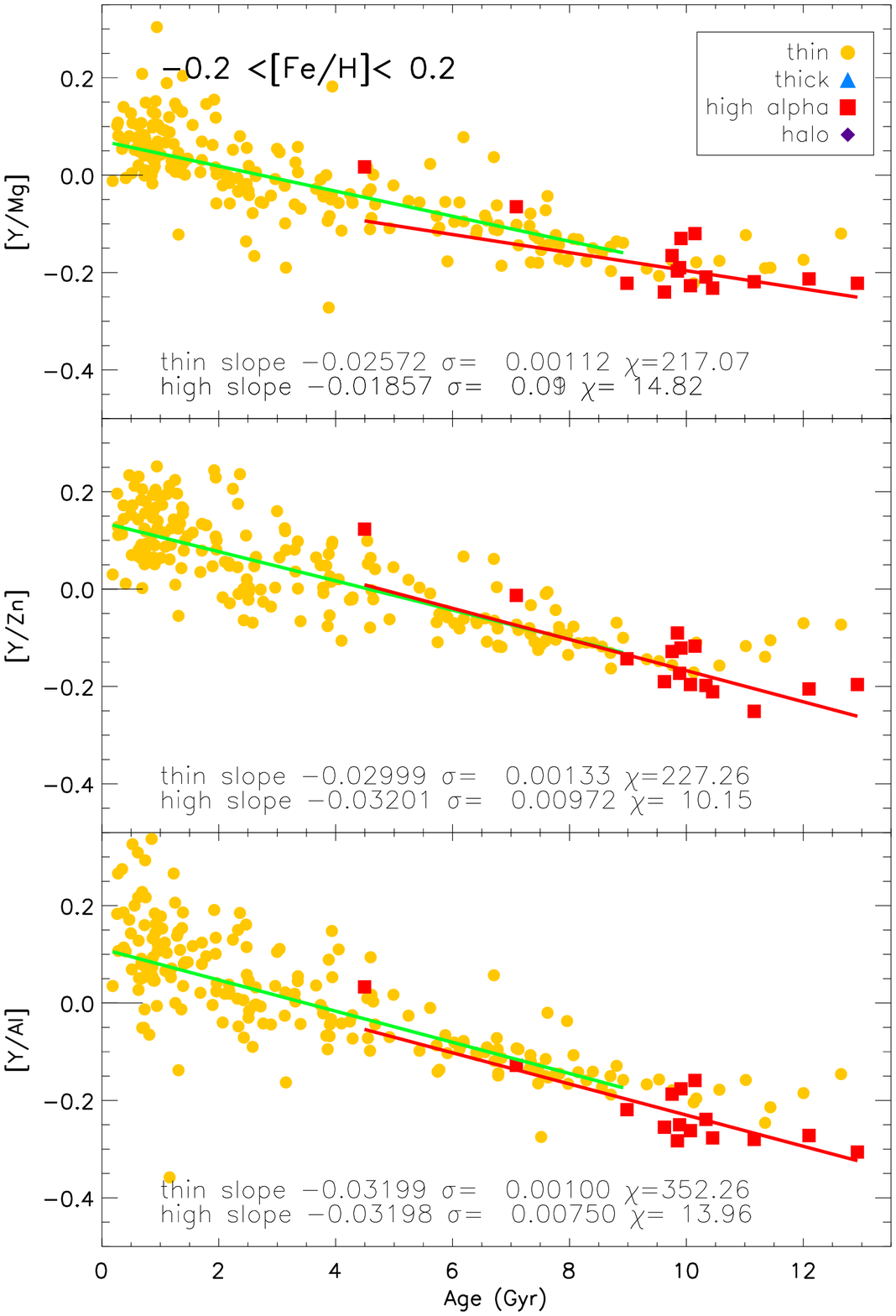}
 \includegraphics[width=6.6cm]{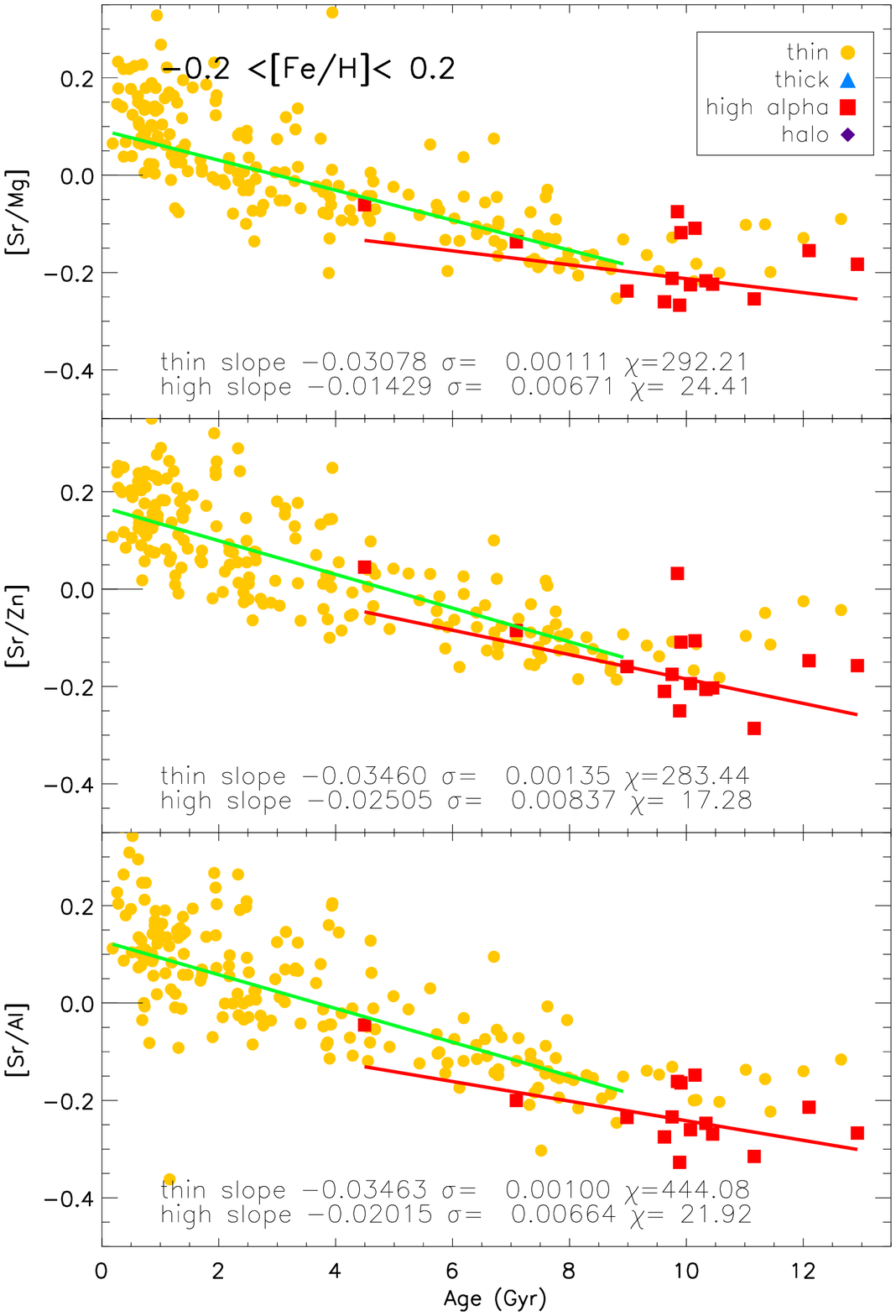}
 \caption{[Y/Mg],[Y/Zn],[Y/Al] and [Sr/Mg],[Sr/Zn],[Sr/Al] for -0.2$<$[Fe/H]$<$0.2. Symbols as in Fig 1.}
   \label{fig3}
\end{center}
\end{figure}

In Fig. \ref{fig1} we can see how several elements depend on age. By combining elements that increase and decrease
with age, respectively, it is possible to have steeper and more constrained trends. For example, [Mg/Fe] shows a tight
increasing trend with age, meanwhile [Eu/Fe], [Zn/Fe] and [Al/Fe] also show this dependency though with more dispersion. 
This trend is expected since these elements are mainly formed by massive stars which started to contribute to the 
Galaxy chemical enrichment earlier than the lower mass stars responsible for Fe production. 
On the other hand, the light-\textit{s} process elements Y and 
Sr show the most clear decreasing trends with age. These elements are formed by low-mass AGB stars so we can expect 
them to increase with time (for younger stars) due to the increasing and delayed contribution of low-mass 
stars as the Galaxy evolves. In Figs. \ref{fig2} and \ref{fig3} we show different combinations of previously mentioned
elements at different metallicity regions. Previous works have explored and confirmed the tight correlation of these 
abundance ratios with age (e.g. \cite{dasilva12}, \cite{nissen15}, \cite{spina16}) but only using solar twins 
or solar analogues. However, \cite{feltzing17} noted that [Y/Mg] clock is not valid at [Fe/H]\,$<$\,-0.5\,dex. 
In our sample we find that still at [Fe/H]\,$<$\,-0.5\,dex the different abundance ratios show a correlation 
with age (steeper for [Sr/X] than for [Y/X]) but this is only valid for thin disk stars. We note however that our 
sample of thick disk stars with reliable ages is quite small. It is also clear that the trends become flat 
at ages $\gtrsim$\,8\,Gyr. On the other hand, at higher metallicities, in the bin -0.2\,$<$\,[Fe/H]\,$<$\,0.2\,dex, 
the abundance ratios of Y and Sr (with respect to Mg, Zn and Al) present similar slopes. Nevertheless, we remark that 
meanwhile [Sr/Fe] presents a constant correlation with age at different metallicities, [Y/Fe] becomes 
flatter as [Fe/H] increases. Moreover, we can observe that thin disk stars present mostly no dependence on age 
for ages $\gtrsim$\,8\,Gyr but \textit{h$\alpha$mr} stars show a continuous dependence in the full age range 
for [Y/X] ratios. The improvement of parallaxes from GAIA DR2 will help to determine more precise ages for our 
stars increasing the sample size and allowing us to better understand the behaviour of the abundance-age trends 
for different populations in the Galaxy.

\begin{acknowledgements}
\begin{scriptsize}
E.D.M., V.Zh.A., N.C.S. and S.G.S. acknowledge the support from Funda\c{c}\~ao para a Ci\^encia 
e a Tecnologia (FCT) through national funds and from FEDER through COMPETE2020 by the following 
grants UID/FIS/04434/2013 \& POCI-01-0145-FEDER-007672, PTDC/FIS-AST/7073/2014 \& 
POCI-01-0145-FEDER-016880 and PTDC/FIS-AST/1526/2014 \& POCI-01-0145-FEDER-016886. 
E.D.M., V.Zh.A., N.C.S. and S.G.S. also acknowledge the support from FCT through Investigador 
FCT contracts IF/00849/2015, IF/00650/2015, IF/00169/2012/CP0150/CT0002 and IF/00028/2014/CP1215/CT0002 
funded by FCT (Portugal) and POPH/FSE (EC). This research has made use of the SIMBAD database 
operated at CDS, Strasbourg (France).

This work has made use of data from the European Space Agency (ESA)
mission {\it Gaia} (https://www.cosmos.esa.int/gaia), processed by
the {\it Gaia} Data Processing and Analysis Consortium (DPAC,
https://www.cosmos.esa.int/web/gaia/dpac/consortium). Funding for
the DPAC has been provided by national institutions, in particular the
institutions participating in the {\it Gaia} Multilateral Agreement.

\end{scriptsize}
\end{acknowledgements}

\end{document}